\documentclass{article}
\usepackage{tabularx, graphicx, booktabs}

\AtBeginDocument{%
  }

\begin{document}

\title{To Vibe Research or Not to Vibe Research? Generative AI in Qualitative Research}

\author{Katja Karhu
\\katja.karhu@lut.fi, 
\\LUT University, Lahti, Finland
\and
Kari Smolander
\\kari.smolander@lut.fi
\\LUT University, Lappeenranta, Finland
\and
Jussi Kasurinen
\\jussi.kasurinen@lut.fi
\\LUT University, Lappeenranta, Finland
}


\maketitle

\begin{abstract}
There has been intense debate among qualitative researchers about whether generative AI is suitable for qualitative research. In this paper, we summarize the broader ongoing discussion of generative AI in qualitative research and its implications for software engineering researchers. The qualitative research approach, small-q (positivist or post-positivist) or Big Q (non-positivist), is among the major criteria for determining whether generative AI can be used in qualitative research. In addition to research philosophy and research approach, skills, ethics, and personal preferences also play a role in researchers' decisions about whether to use AI in qualitative research. 

\end{abstract}





\section{Introduction}


Recently, 419 qualitative researchers rejected the use of generative AI for qualitative research, including high-profile specialists such as Braun and Clarke, the creators of thematic analysis \cite{jowsey_we_2025}. A lively debate on the use of generative AI in qualitative research is ongoing \cite{jowsey_we_2025, friese_beyond_2026, nguyen-trung_documenting_2025}. Some find generative AI useful \cite{bijker_chatgpt_2024, friese_coding_2026}, and some do not \cite{vikan_reflecting_2025, jowsey_frankenstein_2025}, sometimes the results are mixed \cite{de_morais_leca_applications_2025}, and, as mentioned, some reject it completely \cite{jowsey_we_2025, nguyen_engaged_2025}. The question is, why are the researchers' views conflicting? Additionally, why is this debate relevant for software engineering researchers?


There are two reasons why empirical software engineering researchers should also be involved in, or at least follow this discussion on generative AI in qualitative research. First, we use qualitative research methods; second, we develop AI solutions for qualitative research.

Qualitative research in software engineering is important because software development presents a number of unique management and organizational issues that cannot be researched with quantitative methods alone \cite{seaman_qualitative_1999}. Qualitative research methods usually stem from fields outside software engineering, but they are context-independent. For example, grounded theory originated in healthcare research \cite{glaser_discovery_1967}, and thematic analysis originated in psychology \cite{braun_using_2006}, and both are mainstream qualitative research methods among software engineering researchers today. The same good practices in qualitative research in other fields are also good practices in software engineering. Learning from other disciplines is essential for advancing methodologically rigorous and relevant software engineering research \cite{seaman_qualitative_2025}. A wider variety of qualitative methods provides richer, more comprehensive insights in software engineering research \cite{seaman_qualitative_2025}.

Additionally, software engineering researchers are developing new AI systems for qualitative analysis \cite{de_morais_leca_applications_2025}. However, a common criticism among qualitative research experts is that technical people developing AI systems for qualitative research lack knowledge of qualitative research and research philosophy. Nguen and Welch \cite{nguyen_generative_2026} found that the sources claiming benefits of generative AI are often not backed up by qualitative analysis specialists, but the main sources are early adopters (e.g., computer scientists) without qualitative research expertise, major software providers (e.g., NVivo), or recent start-ups developing AI systems for qualitative data analysis. Nguyen-Trung and Friese \cite{nguyen-trung_methodological_2025}, although advocating the use of generative AI in qualitative research, also criticize AI-driven qualitative research studies authored by computer scientists or researchers with a technical background. It should also be noted that Friese represents one of the startups mentioned by Nguyen and Welch \cite{nguyen_generative_2026}. 



Yang et al. \cite{yang_positioning_2026} argue that there is a need for methodological guidance, especially for mitigating AI hallucinations while remaining rigorous and sustainable in software engineering research. In our view, we should also consider the broader research philosophical implications of adopting AI. In this paper, we explore when generative AI can be used in qualitative research, especially in qualitative data analysis. It is worth noting that other potential uses of generative AI in qualitative research, such as conducting interviews or replacing human research participants, have been reported \cite{nguyen_generative_2026}. Our goal is to help software engineering researchers and developers of AI systems for qualitative research make informed decisions about the use of generative AI. We also hope to increase understanding of different research philosophies in qualitative research among software engineering researchers, especially among those who are new to qualitative research. 

\section{Background}

To understand the impact of generative AI on qualitative research, we need to understand the nuances of qualitative research and research philosophy. Qualitative research is not a monolith, as there are different methods and approaches to it. A major factor in determining the suitability of generative AI for qualitative research is the underlying research philosophy or approach.

\subsection{Different Approaches to Qualitative Research}
 
 As stated earlier, over 419 qualitative researchers rejected the use of generative AI, specifically for Big-Q qualitative approaches \cite{jowsey_we_2025}. The keyword here is "Big-Q". The term comes from Kidder and Fine \cite{kidder_qualitative_1987}, who described two very different approaches to qualitative research: small-q and Big-Q. Small-q research draws from the quantitative and post(positivist) research traditions and values consensus, generalizability, and validity \cite{kidder_qualitative_1987, braun_toward_2023}, while Big-Q research is non-positivist and aims for uncovering unusual or surprising phenomena and highlights the researchers' subjectivity as a resource \cite{kidder_qualitative_1987, braun_toward_2023, marecek_numbers_2011}. With small-q research, there is a smaller likelihood that it will uncover surprising or novel phenomena, as a surprising response can be discarded as an outlier \cite{marecek_numbers_2011}. On the other hand, the creativity in Big-Q research can risk eroding rigor in some cases but can also reveal new insights and perspectives beyond familiar templates and checklists \cite{seaman_qualitative_2025}.
 
 We have illustrated some of the key differences between small-q and Big-Q approaches in Table \ref{tab:differences}. The main difference between small-q and Big-Q approaches is the research philosophy: small-q research is positivist or postpositivist, whereas Big-Q research is non-positivist (e.g., interpretivism, constructivism). Positivism or postpositivism aims for objectivity \cite{lee_generalizing_2003}. In positivism, the researcher and the reality are completely separate. The researcher does not influence the studied phenomenon or participants \cite{wohlin_towards_2015, lenberg_qualitative_2024}. Post-positivism differs from positivism by recognizing that although there are universal laws or truths, they can only be imperfectly apprehended by humans \cite{guba_competing_1994}. In addition, post-positivists recognize that research is an interaction between the researcher and the research subject, and that the researchers' values and views influence the inquiry \cite{guba_competing_1994}. This influence is unwanted in post-positivism: researcher bias is seen as a problem and should be mitigated \cite{melegati_surfacing_2021}. 

On the other side, Big-Q research is non-positivist (e.g., constructivist, interpretivist). Overall, the goal of finding universal laws is considered unsuitable for the study of humans because individuals, groups, and other social units are unique and operate in different contexts \cite{lee_generalizing_2003}. The interaction between the researchers and the research subject is vital: knowledge is built through this interaction \cite{melegati_surfacing_2021}. Additionally, the researchers' subjectivity is embraced as a resource \cite{braun_toward_2023}. 



\begin{table*}
    \centering
    \caption{Difference between small-q and Big-Q approaches}

    \begin{tabularx}{\textwidth}{lXX}
        \toprule
         & \textbf{small-q} & \textbf{Big-Q}\\
         \midrule
          Research philosophy & (Post)positivist \cite{braun_toward_2023} & Non-positivist (e.g., interpretivist, constructivist) \cite{braun_toward_2023} \\
          Primary goal & Hypothesis-testing, reaching consensus, discovering universal laws, seeking answers \cite{kidder_qualitative_1987, marecek_numbers_2011, braun_toward_2023, lee_generalizing_2003} & Uncovering surprising, unusual, or novel phenomena, seeking questions \cite{marecek_numbers_2011, kidder_qualitative_1987} \\
        Values & Objectivity as a goal \cite{braun_toward_2023} & Subjectivity as a resource \cite{kidder_qualitative_1987, braun_toward_2023} \\
        Definition of rigor & Replicability, generalizability, validity, intercoder reliability \cite{marecek_numbers_2011, hoda_qualitative_2024, oconnor_intercoder_2020} & Methodological congruence (integrity), reflexive openness (transparency) \cite{braun_reporting_2025}, \\
        Seen as problems & Researcher bias \cite{melegati_surfacing_2021, braun_toward_2023} & "Positivism-creep" \cite{braun_toward_2023} \\
        Data collection & Standardized instrument and setting \cite{marecek_numbers_2011} & Varied instruments and settings \cite{marecek_numbers_2011} \\
        Method examples & Coding reliability thematic analysis \cite{braun_toward_2023}, Glaserian grounded theory \cite{hoda_qualitative_2024}, Socio-technical grounded theory \cite{hoda_qualitative_2024} & Reflexive thematic analysis  \cite{braun_toward_2023}, Strauss-Corbinian grounded theory  \cite{hoda_qualitative_2024}, Constructivist grounded theory \cite{hoda_qualitative_2024}, Socio-technical grounded theory \cite{hoda_qualitative_2024}\\

        \bottomrule
    \end{tabularx}
    \label{tab:differences}
\end{table*}

In small-q research, concepts from quantitative research, such as replicability, generalizability, and validity, are important for evaluating research quality \cite{marecek_numbers_2011, hoda_qualitative_2024}. But if these same concepts are applied in Big-Q research, it results in "positivism-creep", which is an example of methodological incongruence \cite{braun_reporting_2025, braun_toward_2023}. Methodological incongruence means that the research question, chosen methodology (including sampling, data collection, and data analysis), findings, and researcher´s philosophical perspective are not aligned \cite{willgens_how_2016}.  In Big-Q research, indicators of high-quality research are methodological congruence (integrity) and reflexive openness (transparency) \cite{braun_reporting_2025}. They capture what the researcher did and how they did it: theoretical assumptions, research goals, practices of generating evidence, analytical processes, researcher positionality and subjectivity, and openness with research participants \cite{braun_reporting_2025}.  


 Table \ref{tab:differences} also contains examples of commonly used qualitative methods in software engineering from both the small-q and Big-Q categories. Note that this listing includes only a small subset of all qualitative research methods. Some methods can fall into either category, depending on the philosophical stance taken in a study. Socio-technical grounded theory is one such example; studies can be conducted using either a positivist or a non-positivist stance \cite{hoda_qualitative_2024}. Thematic analysis, as a family of methods, also includes small-q and Big-Q methods \cite{braun_toward_2023}.

\subsection{Understanding of Qualitative Research in Software Engineering}

One of the major criticisms of studies on generative AI in qualitative research is that they lack expertise in qualitative research \cite{nguyen_generative_2026, nguyen-trung_methodological_2025}. A concrete example of this is attempting to validate AI's performance by calculating inter-rater reliability, when the qualitative research framework, such as reflexive thematic analysis, rejects inter-rater reliability as a metric because coding is inherently subjective and interpretive. \cite{nguyen-trung_methodological_2025}. Using inter-rater reliability, a small-q concept, to evaluate a Big-Q process is a clear example of methodological incongruence \cite{nguyen-trung_methodological_2025}. This is criticism we, as software engineers, should take into account: we need input and guidance from qualitative research experts when evaluating or developing AI systems for qualitative analysis.

 A lack of training or knowledge of research philosophy is also a known issue within the software engineering research community  \cite{hoda_qualitative_2024, melegati_surfacing_2021}. So far, this problem has mostly arisen in the peer review process. When reviewers and editors are unfamiliar with different approaches to qualitative research, they can accidentally mix qualitative approaches, threatening the credibility of methods and the applicability of outcomes \cite{hoda_qualitative_2024}; compromising the methodological integrity of qualitative research \cite{clarke_being_2025}. A reviewer aligned with a (post)positivist (Big-Q) stance may expect all qualitative studies they review to demonstrate quality criteria such as generalisability, reproducibility, and replicability \cite{hoda_qualitative_2024}. And vice versa, a non-positivist reviewer may give harsh reviews to (post)positivist (small-q) research \cite{hoda_qualitative_2024}. The reason qualitative research reviews fail is not only the fault of reviewers: authors often do not explicitly state, or even know, their philosophical stance \cite{melegati_surfacing_2021, hoda_qualitative_2024}. Finding and stating the philosophical stance enhances transparency, credibility, and trust in qualitative work \cite{seaman_qualitative_2025}.
 
Historically, in software engineering, especially among researchers with a computer science or formal methods background, qualitative research is often viewed through a small-q lens. This is because of the strong quantitative tradition in software engineering \cite{lee_information_1997, seaman_qualitative_1999}. Therefore, although the small-q and Big-Q division of qualitative research did not originate in the software engineering context, it describes the "opposing forces" at play in qualitative research in software engineering. For this reason, the small-q/Big Q division can offer a framework for discussing and increasing common understanding of their philosophical stance and different types of qualitative research for software engineering researchers.

\section{When to Use Generative AI in Qualitative Data Analysis}

In general, qualitative analysis is more labor-intensive and time-consuming than quantitative analysis \cite{seaman_qualitative_1999}. The typical expected benefits of using generative AI in qualitative research are efficiency and productivity \cite{de_morais_leca_applications_2025, foley_it_2025, nguyen_generative_2026}. Productivity and efficiency are understandable goals in the era of "publish or perish", where qualitative researchers feel the pressure to publish more \cite{foley_it_2025}. On the other hand, reviewers dread the increase in publication numbers \cite{yang_positioning_2026}. There are also doubts about whether the actual benefits of generative AI in qualitative research are relevant or have been exaggerated, especially in the context of Big-Q research \cite{nguyen_generative_2026}. 

There were three reasons for qualitative researchers rejecting the use of generative AI in Big Q qualitative research \cite{jowsey_we_2025}. First, generative AI is incapable of meaning-making; it generates data based on statistical algorithms and, therefore, cannot be reflexive \cite{jowsey_we_2025}. Second, qualitative research is a distinctly human practice: undertaken by humans, with or about humans, to understand people and social processes  \cite{jowsey_we_2025}. The third reason was the harm caused by big AI corporations, especially environmental harm and exploitation of workers in the Global South \cite{jowsey_we_2025}.


\begin{figure}
    \centering
    \includegraphics[width=1\linewidth]{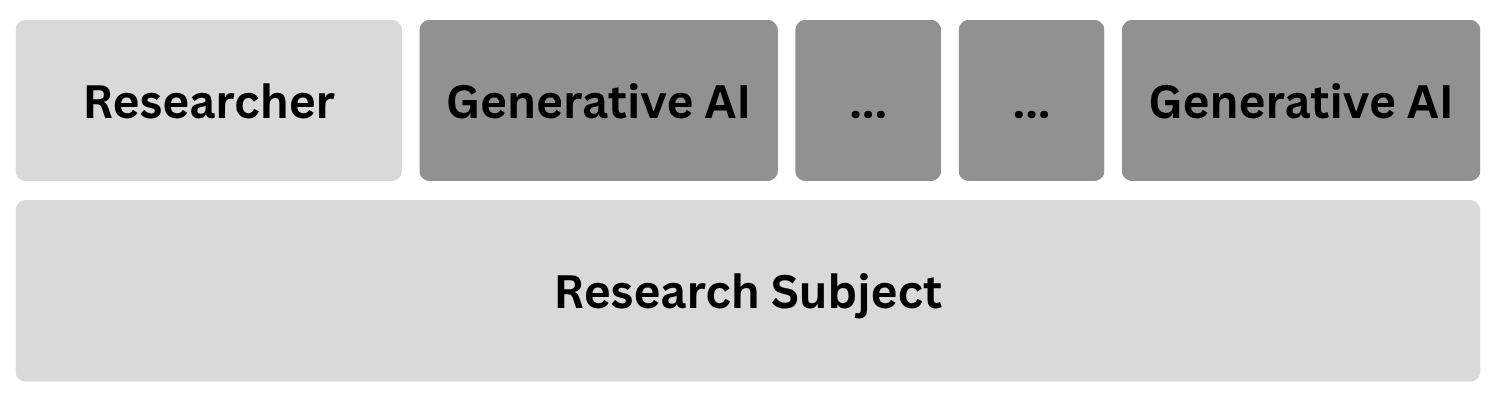}
    \caption{AI as a parallel researcher}
    \label{fig:parallel}
\end{figure}

\begin{figure}
    \centering
    \includegraphics[width=1\linewidth]{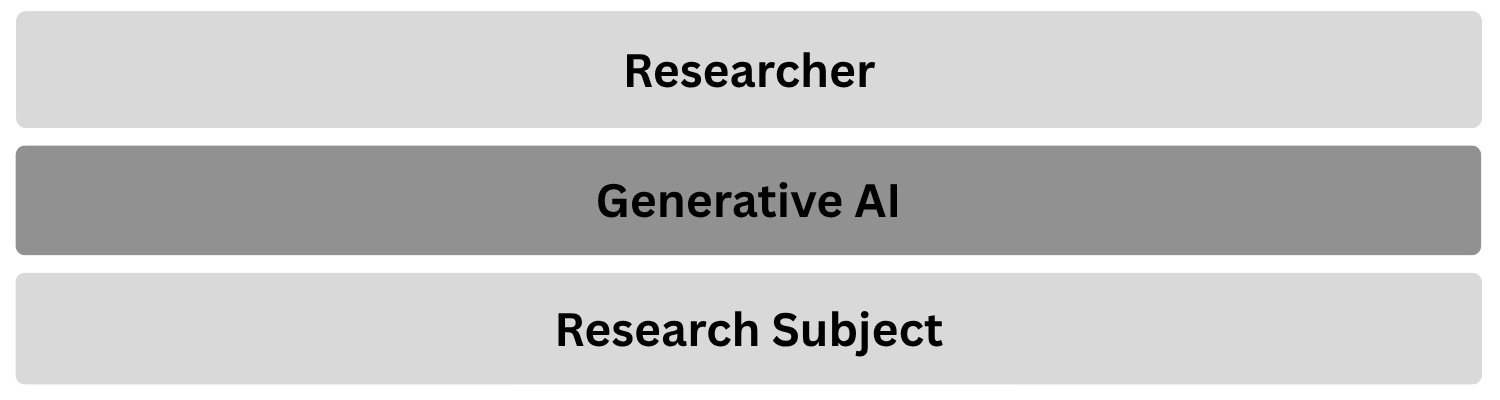}
    \caption{AI as a layer between researcher and subject}
    \label{fig:layer}
\end{figure}

Another important part of the ongoing discussion among qualitative researchers is how AI should be used in qualitative data analysis \cite{friese_beyond_2026}. Should it be used to automate qualitative coding, replace the researcher completely, be a research assistant, or should the whole qualitative analysis process be rethought in order to accommodate AI-based solutions? 

Figures \ref{fig:parallel} and \ref{fig:layer} illustrate, in a simplified way, how generative AI can be perceived in the qualitative research process. There are at least two options. Generative AI could be seen as a parallel system or alternative to a researcher (Figure \ref{fig:parallel}). This would be the case where generative AI would act as an independent coder or analyst, without directly affecting the interaction between the human researcher and the research subject. Alternatively, generative AI can be seen as a layer in the interaction between the researcher and the research subject (Figure \ref{fig:layer}). Examples of this would be generative AI conducting interviews, as well as conversational data analysis approaches proposed by Friese \cite{friese_coding_2026} and Nguyen-Trung \cite{nguyen-trung_chatgpt_2025}, in which AI serves as the interface to research data.

The core of generative AIs and LLMs is statistical pattern recognition, which aligns well with the goals and quantitative traditions of small-q research. Generative AI as a paraller source for analysis (see Figure \ref{fig:parallel} is entirely possible in the small-q context: bias and hallucination can be mitigated, for example, by measuring intercoder reliability. A single model or multiple different generative AI models could be utilized as data "analysts". The human researcher can be a reference point for intercoder reliability by coding and analyzing the data independently. Alternatively, leaving the human researcher out of the analysis could be possible if multiple generative AI models are used and their intercoder reliability is high enough. In post-positivist small-q research, this means that human bias has been eliminated, and only generative AI bias needs to be mitigated. There is potential for faster analysis in small-q analysis. However, as with small-q research in general, there is a risk that more nuanced information might be lost. 


In Big-Q research, where knowledge is built in interaction between the researcher and the research subject, neither of these options (Figures \ref{fig:parallel} and \ref{fig:layer} ) seems suitable. Generative AI would be a problematic additional researcher because it is very difficult to reflect on the AI's background: what training data was used, and how this affects interpretation and analysis. Can AI utilize subjectivity as a resource? And placing AI as an interpreter between the researcher and the research subject \ref{fig:layer} can interfere with the interaction and, therefore, with knowledge creation.

Even if we put aside the philosophical aspects of research, using generative AI in qualitative research, whether small-q or Big-Q, requires AI and technical skills. Friese \cite{friese_coding_2026} has found that general-purpose LLMs are not well-suited for qualitative research: dedicated models, RAG techniques, and fine-tuning are needed to achieve better results \cite{friese_coding_2026, friese_beyond_2026}.
Friese \cite{friese_coding_2026} argues that RAG systems offer more reliability, as the model’s outputs are explicitly grounded in specific passages retrieved from the user’s own data, reducing the risk of hallucination and enabling researchers to validate how AI responses were generated from the source data.

This means the researcher should be well-versed in AI models or willing to invest time in acquiring the skillset needed to work with generative AI. Acquiring the needed technical skills is often a problem for qualitative research specialists \cite{friese_coding_2026}. Overall, engaging in AI-assisted qualitative research is demanding, as it requires both qualitative research and AI skills. 
Therefore, more collaboration between technical people and qualitative research experts is needed.

Because there is no proof yet that generative AI can conduct superior qualitative analysis when compared to humans, especially in Big-Q research, there is still room for human researchers and personal preference. Being the human-in-the-loop, or acting as a (micro)manager for AI tools, is not appealing to all qualitative researchers, including the authors of this paper. For many, there is joy in finding codes and their connections and building themes from the ground up \cite{levitt_consideration_2026}.

On the other hand, those who find qualitative coding unappealing and cumbersome might find more joy in working with new technologies, such as generative AI. For them, small-q research would be a safer option.

 As researchers have an incentive to publish more, we may see qualitative research skewing toward small-q research, as it might be faster to conduct with AI assistance. In general, Traberg et al. \cite{traberg_ai_2026} have raised concerns that AI is turning research into a scientific monoculture and highlighted the critical role of researchers trained in non-computational traditions in sustaining methodological diversity. De Morais Leça \cite{de_morais_leca_applications_2025} have identified a similar issue in the context of software engineering research: while some researchers view generative AI as a means to reduce human bias, the composition of the underlying data may favor dominant perspectives, potentially standardizing scientific inquiry. 

There are also valid reasons for rejecting the use of generative AI, such as ethical concerns. Jowsey et al. \cite{jowsey_we_2025} specifically mention environmental impact and the treatment of workers in the Global South as among the primary reasons for rejecting the use of generative AI. Other ethical concerns include issues such as data privacy and transparency in research \cite{de_morais_leca_applications_2025}. 

Miceli et al. \cite{miceli_methodological_2025} point out the highly problematic nature of opaque supply chains in AI, ranging from the mining of rare-earth minerals to the exploitation of data workers. Data workers face issues, such as exposure to psychologically disturbing content, low wages, wage theft, and exploitation of people in vulnerable positions \cite{miceli_methodological_2025}. They have found that these issues are underresearched and highlight the need to examine the human costs of AI infrastructure and to center the voices of the labor behind AI \cite{miceli_methodological_2025}. There is more room for research in sustainable and ethical AI, also in software engineering.

On the other hand, as Wheeler \cite{wheeler_technological_2026} states, placing the responsibility for resisting exploitative systems on individual researchers is insufficient when generative AI is governed by a small number of corporate actors whose commercial interests define what research should be done and is preferred. Governance and regulation of the exploitative systems, as well as ethical alternatives to big AI corporations, are needed.


\section{Conclusion and Reflections}



Unless you are highly knowledgeable in the philosophy of science, a leading expert in qualitative research, and an expert in generative AI, we do not recommend engaging in Big-Q qualitative data analysis using generative AI. Our focus in this paper was specifically on the qualitative data analysis. There are also other aspects of using generative AI in qualitative research, or "vibe researching", such as brainstorming, that warrant evaluation from a research-philosophical lens. It is possible that AI assistance may help Big-Q research in some way, but it is more likely that generative AI solutions are not essential for understanding humans and human organizations, especially when using subjectivity as a research resource. 

On the other hand, for small-q research, generative AI can be useful. For example, hallucinations can be managed by using quantitative metrics, such as intercoder reliability, to evaluate AI performance. There is, however, a risk that qualitative research will skew towards small-q research if it is faster to conduct with AI assistance. 

As the authors of this paper, we believe that learning about and understanding the object of research through human-driven data collection and qualitative analysis is essential to Big-Q research. It is very likely that, by reading AI output, we do not achieve the same level of learning and understanding as we do with direct interaction with the research subject and the data.

The question of whether to use generative AI in qualitative research is deeply epistemological and rooted in research philosophy. Is the purpose of research to produce texts quickly, or is the purpose to create a deeper understanding of the world? If our purpose is to produce texts for publication, then AI is a perfect tool. It can produce plausible-looking texts and classifications with great efficiency. On the other hand, if our purpose is to improve our understanding, explanation, and interpretation of the world and improve the wealth and lives of the human species, then AI may have its uses, but it is the responsibility of human beings to understand, explain, and interpret. We wish that AI in qualitative research would not weaken humans' ability to understand, explain, and interpret without AI's help.




\bibliographystyle{acm}
\bibliography{references}

\end{document}